\documentclass[aps,pre,twocolumn,showpacs,superscriptaddress]{revtex4-1}

\usepackage[final]{graphicx}
\usepackage{amsmath}
\usepackage{verbatim}
\usepackage{color}
\usepackage{ulem}

\usepackage{threeparttable}

\usepackage{multirow}
\usepackage[colorlinks,
linkcolor=blue,
citecolor=blue,
urlcolor=blue]{hyperref}

\begin{document}

% Use the \preprint command to place your local institutional report
% number in the upper righthand corner of the title page in preprint mode.
% Multiple \preprint commands are allowed.
% Use the 'preprintnumbers' class option to override journal defaults
% to display numbers if necessary
\preprint{}

%Title of paper
\title{Burst behavior due to quasimode excited by stimulated Brillouin scattering in high-intensity laser-plasma interaction}
% repeat the \author .. \affiliation  etc. as needed
% \email, \thanks, \homepage, \altaffiliation all apply to the current
% author. Explanatory text should go in the []'s, actual e-mail
% address or url should go in the {}'s for \email and \homepage.
% Please use the appropriate macro foreach each type of information

% \affiliation command applies to all authors since the last
% \affiliation command. The \affiliation command should follow the
% other information
%\author{}
% \affiliation can be followed by \email, \homepage, \thanks as well.
%\email[]{chengzhuo@pku.edu.cn}
%\homepage[]{Your web page}
%\thanks{}
%\altaffiliation{}
\author{Q. S. Feng} 
\affiliation{Institute of Applied Physics and Computational
	Mathematics, Beijing, 100094, China}

\author{L. H. Cao} 
\email{cao\_lihua@iapcm.ac.cn}
\affiliation{Institute of Applied Physics and Computational
	Mathematics, Beijing, 100094, China}
\affiliation{HEDPS, Center for
	Applied Physics and Technology, Peking University, Beijing 100871, China}
\affiliation{Collaborative Innovation Center of IFSA (CICIFSA) , Shanghai Jiao Tong University, Shanghai, 200240, China}

\author{Z. J. Liu} 
\affiliation{Institute of Applied Physics and Computational
	Mathematics, Beijing, 100094, China}
\affiliation{HEDPS, Center for
	Applied Physics and Technology, Peking University, Beijing 100871, China}

\author{C. Y. Zheng} 
 \email{zheng\_chunyang@iapcm.ac.cn}
\affiliation{Institute of Applied Physics and Computational
	Mathematics, Beijing, 100094, China}
\affiliation{HEDPS, Center for
	Applied Physics and Technology, Peking University, Beijing 100871, China}
\affiliation{Collaborative Innovation Center of IFSA (CICIFSA) , Shanghai Jiao Tong University, Shanghai, 200240, China}

\author{X. T. He} 
\affiliation{Institute of Applied Physics and Computational
	Mathematics, Beijing, 100094, China}
\affiliation{HEDPS, Center for
	Applied Physics and Technology, Peking University, Beijing 100871, China}
\affiliation{Collaborative Innovation Center of IFSA (CICIFSA) , Shanghai Jiao Tong University, Shanghai, 200240, China}

%\address[address1]{HEDPS, Center for
%	Applied Physics and Technology, Peking University, Beijing 100871, China}
%\address[address2]{Institute of Applied Physics and Computational
%	Mathematics, Beijing, 100094, China}
%\address[address3]{Collaborative Innovation Center of IFSA (CICIFSA) , Shanghai Jiao Tong University, Shanghai, 200240, China}
%Collaboration name if desired (requires use of superscriptaddress
%option in \documentclass). \noaffiliation is required (may also be
%used with the \author command).
%\collaboration can be followed by \email, \homepage, \thanks as well.
%\collaboration{}
%\noaffiliation

\date{\today}

\begin{abstract}
The strong-coupling mode, called \textquotedblleft quasimode\textquotedblright, will be excited by stimulated Brillouin scattering (SBS) in high-intensity laser-plasma interaction. And SBS of quasimode will compete with SBS of fast mode (or slow mode) in multi-ion species plasmas, thus leading to a low-frequency burst behavior of SBS reflectivity. The competition of quasimode and ion-acoustic wave (IAW) is an important saturation mechanism of SBS in high-intensity laser-plasma interaction. These results give a clear explanation to the low-frequency periodic burst behavior of SBS and should be considered as a saturation mechanism of SBS in high-intensity laser-plasma interaction.
	
\end{abstract}

% insert suggested PACS numbers in braces on next line
%\pacs{52.38.Bv, 52.35.Fp, 52.35.Mw, 52.35.Sb}
% insert suggested keywords - APS authors don't need to do this
%\keywords{}

%\maketitle must follow title, authors, abstract, \pacs, and \keywords
\maketitle

% body of paper here - Use proper section commands
% References should be done using the \cite, \ref, and \label 
%\section{Introduction}
%\section{\label{Sec: Introduction}Introduction}
Backward stimulated Brillouin scattering (SBS), a three-wave interaction process where an incident electromagnetic wave (EMW) decays into a backscattered EMW and a forward propagating ion-acoustic wave (IAW), leads to a great energy loss of the incident laser and is detrimental in inertial confinement fusion (ICF) \cite{He_2016POP,Glenzer_2010Science,Glenzer_2007Nature}. Therefore, SBS plays an important role in the successful ignition goal of ICF. 
Multiple ion species are contained in the laser fusion program \cite{Neumayer_2008PRL}. In the indirect-drive ICF \cite{Glenzer_2010Science,Glenzer_2007Nature} or the hybrid-drive ignition \cite{He_2016POP}, the inside of hohlraum will be filled with low-Z plasmas, such as H or CH plasmas from the initial filled material or from the ablated material off the capsule. Typically, in hybrid-drive ICF, the laser intensity can reach as high as $I_0\sim10^{16}W/cm^2$, the strong-coupling modes will be excited by SBS. The stimulated scattering process in this regime is referred to as SBS in the strong coupling regime or in the quasimode regime. \cite{LiuCS_1974POF,Guzdar_1996POP} The quasimode or the strong-coupling mode refers to the modified low-frequency mode involved in the three-wave process, while the electrostatic mode refers to a natural mode of the system in the absence of the pump wave. The burst behavior of SBS reflectivity is universal in high-intensity laser-plasma interaction, however, the cause is not clear and require the explanation. Besides, understanding the excitation of different IAW modes  and competition between SBS of different modes under the condition of high-intensity laser plasma interaction is vital to predict SBS laser energy losses and to improve the energy coupling into the fusion capsule. 

Many mechanisms for the saturation of SBS have been proposed, including frequency detuning due to particle trapping \cite{Froula_2002PRL}, coupling with higher harmonics \cite{Bruce_1997POP, Rozmus_1992POP}, increasing linear Landau damping by kinetic ion heating \cite{Rambo_1997PRL, Pawley_1982PRL}, the creation of cavitons in plasmas \cite{ Weber_2005PRL, Weber_2005POP} and so on. However, the burst behavior of SBS in high-intensity laser-plasma interaction is confusing and has not been explained well, which may be a potential saturation mechanism of SBS.

 In this paper, we report the first demonstration that the strong-coupling mode or quasimode will be excited by SBS and will coexist and compete with the IAW in the high-intensity laser-plasma interaction. The competition between SBS of IAW and SBS of quasimode will lead to the low-frequency burst behavior of SBS reflectivity and decrease the total SBS reflectivity. Therefore the competition of quasimode and IAW  excited by SBS is an important saturation mechanism of SBS in high-intensity laser-plasma interaction.

%\section{\label{Sec:Theory analysis}Theoretical analysis}

The wave number of IAW excited by backward SBS can be calculated by
\begin{equation}
\label{Eq:k_A}
k_A\lambda_{De}\simeq2\frac{v_{te}}{c}\sqrt{n_c/n_e-1},
\end{equation}
where $v_{te}=\sqrt{T_e/m_e}$ is the electron thermal velocity, $n_e, T_e, m_e$ is the density, temperature and mass of the electron.
Considering fully ionized, neutral, unmagnetized plasmas with the same temperature of all ion species ($T_H=T_C=T_i$), the linear dispersion relation of the ion acoustic wave in multi-ion species plasmas is given by \cite{Williams_1995POP,Feng_2016POP,Feng_2016PRE}
\begin{equation}
\label{Eq:Dispersion}
\epsilon(\omega,k=k_A)=1+\chi_e+\sum_\beta \chi_{i\beta}=0,
\end{equation}
where $\chi_e$ is the susceptibility of electron and $\chi_{i\beta}$ is the susceptibility of ion $\beta$. 

When the strong pump light interacts with plasmas, the strong-coupling mode will be generated, which will grow with time but not be damped and is called quasimode. The dispersion relation of quasimode in strong pump light is given by \cite{Drake_1974POF, Liu_2012CPB}
 \begin{equation}
 \label{Eq:Dispersion of quasimode}
 \epsilon(\omega_A,k_A)=\frac{k_A^2v_{os}^2}{4}\chi_e(1+\sum_\beta \chi_{i\beta})(\frac{1}{D^-}+\frac{1}{D^+}),
 \end{equation}
 where $D^-\equiv D(\omega_A-\omega_0,k_A-k_0)$, $D^+\equiv D(\omega_A+\omega_0,k_A+k_0)$, and $D(\omega, k)=k^2c^2+\omega_{pe}^2-\omega^2$.

\begin{figure}[!tp]
	\includegraphics[width=1.0\columnwidth]{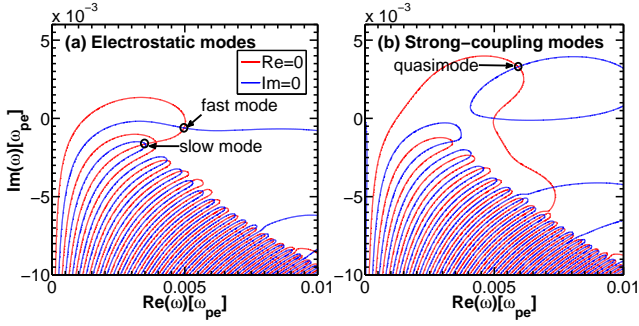}
	
	\caption{\label{Fig:Dispersion}(Color online) Contours of solutions to the dispersion relations of (a) the fast IAW mode and the slow IAW mode without pump light, (b) the quasimode with strong pump light $I_0=1\times10^{16}W/cm^2$. Where the red line is $Re[\epsilon]=0$ and the blue line is $Im[\epsilon]=0$. The conditions are: $T_e=5keV$, $T_i=0.2T_e$, $n_e=0.3n_c$ and $k_{A}\lambda_{De}=0.3$ in C$_2$H plasmas.}
\end{figure}

Under the condition of $T_e=5keV$, $n_e=0.3n_c$, one can obtain the wave number of the IAW $k_A\lambda_{De}=0.3$ from Eq. (\ref{Eq:k_A}). By solving Eq. (\ref{Eq:Dispersion}) and Eq. (\ref{Eq:Dispersion of quasimode}), the contours of solutions of the IAWs without pump light and with strong pump light are shown in Fig. \ref{Fig:Dispersion} under the condition of $T_i=0.2T_e$, $k_A\lambda_{De}=0.3$. Although there are infinite solutions of the IAW as shown in Fig. \ref{Fig:Dispersion}, the mode with the least Landau damping ($|Im(\omega)|$) will be preferentially excited in SBS. There exist two groups of modes called \textquotedblleft the fast mode" and \textquotedblleft the slow mode" in multi-ion species plasmas as shown in Fig. \ref{Fig:Dispersion}(a). Here, the fast mode and the slow mode refer to the least damped mode belonging to each class of modes. When strong pump light interacts with plasmas, the quasimode will be excited. As shown in Fig. \ref{Fig:Dispersion}(b), the imaginary part of quasimode frequency is positive, which illustrates that the quasimode is a growing mode.

%\section{\label{Sec:Vlasov simulation}Numerical simulation}
An one dimension in space and one dimension in velocity (1D1V) Vlasov-Maxwell code \cite{Liu_2009POP} is used to research the quasimode excited by SBS in multi-ion species plasmas. We choose the high-temperature and high-density region as an example, the electron temperature and electron density are $T_e=5keV$, $n_e=0.3n_c$, where $n_c$ is the critical density for the incident laser. The electron density is taken to be higher than $0.25n_c$, thus the stimulated Raman scattering \cite{Feng_2018POP} and two-plasmon decay instability \cite{Xiao_2015POP,Xiao_2016POP} are excluded. The C, C$_2$H, CH, H plasmas are taken as typical examples since they are common in ICF \cite{He_2016POP,Glenzer_2007Nature}. The ion temperature is $T_i=0.2T_e$. The linearly polarized laser intensity is $I_0=1\times10^{16}W/cm^2$ with wavelength $\lambda_0=0.351\mu m$. The spatial scale is [0, $L_x$] discretized with $N_x=5000$ spatial grid points and spatial step $dx=0.2c/\omega_0$. And the spatial length is $L_x=1000c/\omega_0\simeq160\lambda_0$ with $2\times5\%L_x$ vacuum layers and $2\times5\%L_x$ collision layers in the two sides of plasmas boundaries. The plasmas located at the center with density scale length $L=0.8L_x$ are collisionless. The incident laser propagates along the $x$ axis from the left to the right with outgoing boundary conditions. The strong collision damping layers are added into the two sides of the plasmas boundaries ($2\times5\%L_x$) to damp the electrostatic waves such as IAWs at the boundaries and decrease the effect of sheath field. The electron velocity scale $[-0.8c, 0.8c]$ and the ion velocity scale $[-0.03c, 0.03c]$ are discretized with $2N_v+1$ ($N_v=512$) grid points. The total simulation time is $t_{end}=1\times10^5\omega_0^{-1}$ discretized with $N_t=5\times10^5$ and time step $dt=0.2\omega_0^{-1}$.
  \begin{figure}[!tp]
  	%	\centering
  	\includegraphics[width=1\columnwidth]{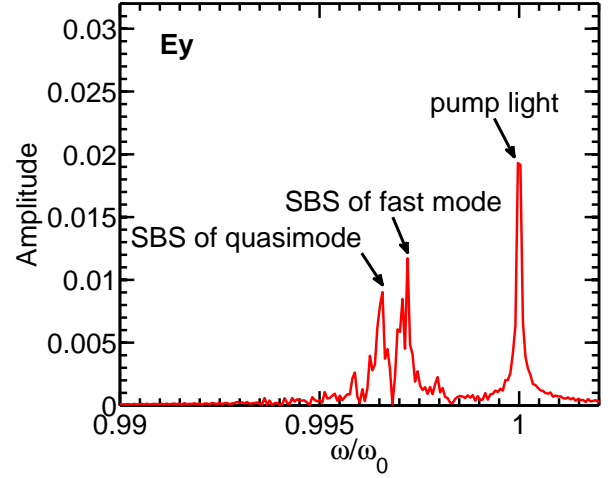}
  	\caption{\label{Fig:Spectra_w}(Color online) The frequency spectra of $Ey$ with the time scope $t\in[0, 1\times10^5]\omega_0^{-1}$ at $x_0=25c/\omega_0$. 
  		The parameters are $n_e=0.3n_c, T_e=5keV, T_i=0.2T_e, I_0=1\times10^{16}W/cm^2$ in C$_2$H plasmas as the same as Fig. \ref{Fig:Dispersion}(b). }
  \end{figure}
  
   \begin{table}
   	\caption{\label{table1} The frequencies of different modes and corresponding scattering lights. The conditions are: $T_e=5keV$, $T_i=0.2T_e$, $n_e=0.3n_c$, $k_{A}\lambda_{De}=0.3$, and $I_0=1\times10^{16}W/cm^2$ in C$_2$H plasmas. }
   	
   	%\begin{threeparttable}
   	\begin{ruledtabular}
   		
   		\begin{tabular}{cccc}
   			\hline
   			\multirow{2}*{Mode}& \multicolumn{2}{|c|}{Theory} & Simulation\\
   			\cline{2-4}
   			&\multicolumn{1}{|c}{$\omega_A/[10^{-3}\omega_0]$}&\multicolumn{1}{c|}{$\omega_s/\omega_0$} & $\omega_s/\omega_0$
   			\\
   			\cline{1-4}
   			\multicolumn{1}{c}{fast mode} & \multicolumn{1}{|c}{2.7} & \multicolumn{1}{c|}{0.9973} & 0.9972
   			\\
   			\cline{1-4}
   			\multicolumn{1}{c}{slow mode} & \multicolumn{1}{|c}{1.9} & \multicolumn{1}{c|}{0.9981} & $\setminus$
   			\\
   			\cline{1-4}   			
   			\multicolumn{1}{c}{quasimode} & \multicolumn{1}{|c}{3.8} & \multicolumn{1}{c|}{0.9962} & 0.9966
   			\\
   			\cline{1-4}   			
   			
   		\end{tabular}
   		
   	\end{ruledtabular}
   	
   	%\end{threeparttable}%
   \end{table}
   
 Figure \ref{Fig:Spectra_w} shows the spectra of the SBS scattering lights in the case of C$_2$H plasmas. Calculated by Vlasov simulation, the two peak frequencies of scattering lights in C$_2$H plasmas are $\omega_s=0.9972\omega_0, 0.9966\omega_0$. Calculated from Eq. (\ref{Eq:Dispersion}) and Eq. (\ref{Eq:Dispersion of quasimode}), the frequency of the fast mode is $\omega_f=4.98\times10^{-3}\omega_{pe}=2.73\times10^{-3}\omega_0$, the frequency of the slow mode is $\omega_s=3.49\times10^{-3}\omega_{pe}=1.9\times10^{-3}\omega_0$ and the frequency of the quasimode with pump light $I_0=1\times10^{16}W/cm^2$ is $\omega_q=6.98\times10^{-3}\omega_{pe}=3.82\times10^{-3}\omega_0$. Therefore, the corresponding theoretical frequencies of SBS scattering lights of the fast mode, the slow mode and the quasimode are $\omega_s=0.9973\omega_0, 0.9981\omega_0, 0.9962\omega_0$. The theoretical and simulation results are shown in Tab. \ref{table1}. Compared to the theoretical frequencies, we can see that the two peaks of the SBS scattering lights in Fig. \ref{Fig:Spectra_w} are indeed the SBS of fast mode and SBS of quasimode.

  \begin{figure}[!tp]
  	\includegraphics[width=1\columnwidth]{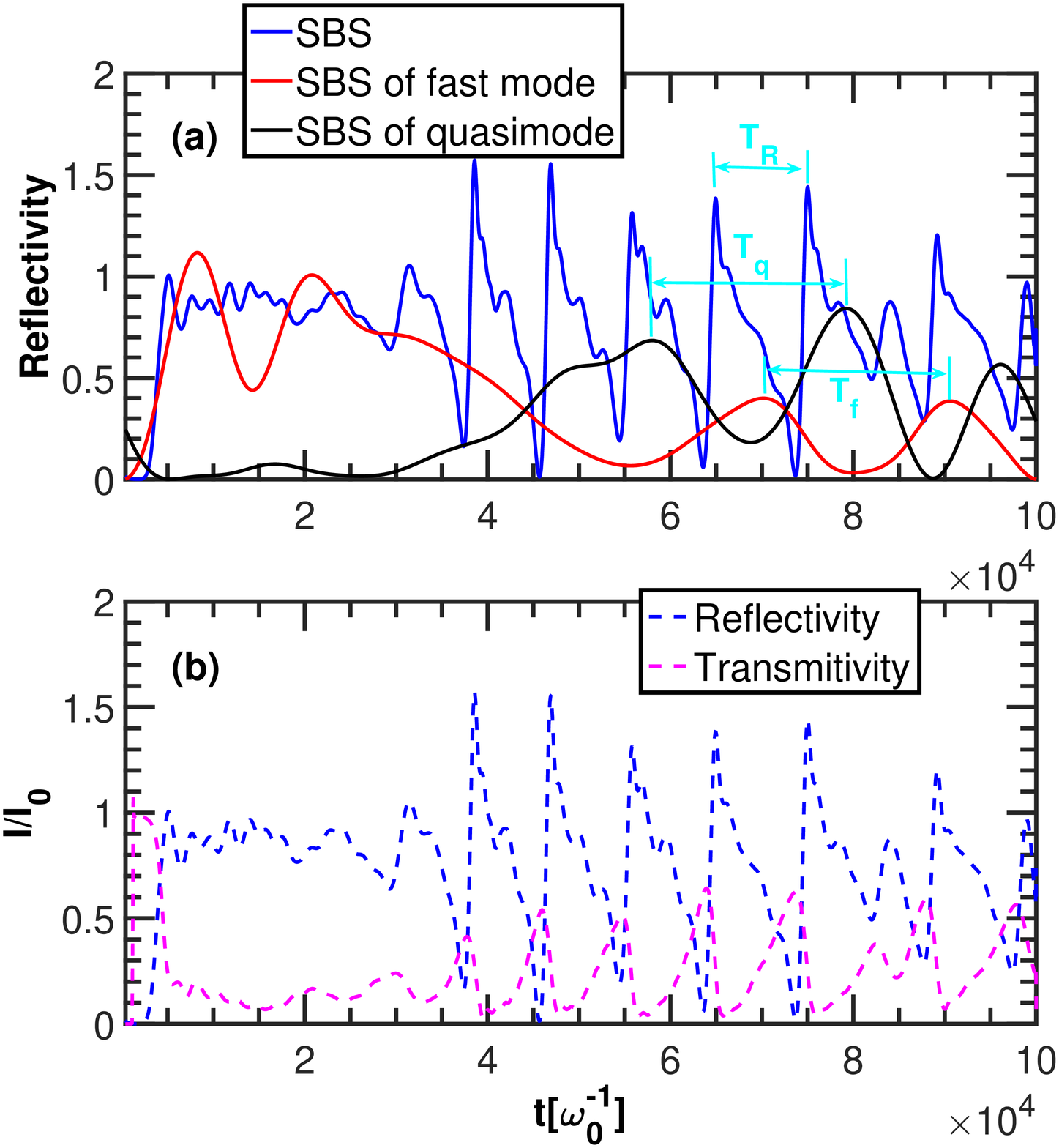}
  	
  	\caption{\label{Fig:Reflectivity}(Color online) (a) The SBS reflectivities of different modes evolve with time. Where SBS is the total SBS with the frequency scope $\omega\in[0.9\omega_0, 0.999\omega_0]$, SBS of fast mode with $\omega\in[0.9968\omega_0, 0.9977\omega_0]$ and SBS of quasimode with $\omega\in[0.9960\omega_0, 0.9968\omega_0]$. (b) The reflectivity and the transmitivity of the total SBS. The condition is as the same as Fig. \ref{Fig:Spectra_w}.}
  \end{figure}
  
 Figure \ref{Fig:Reflectivity} gives the SBS reflectivities of fast mode and quasimode evolution with time. As shown in Fig. \ref{Fig:Reflectivity}(a), the SBS of fast mode will develop quickly before $\sim1\times10^4\omega_0^{-1}$ and saturate during $t\in[1\times10^4, 3\times10^4]\omega_0^{-1}$. In this stage, the total SBS reflectivity (blue line in Fig. \ref{Fig:Reflectivity}(a)) is from the SBS of fast mode. After $3\times10^4\omega_0^{-1}$, SBS of quasimode will develop and compete with SBS of fast mode, which will lead to the burst behavior of the total SBS reflectivity. The SBS of fast mode demonstrates an obvious decrease when the SBS of quasimode develops to the peak amplitude during $[3\times10^4, 6\times10^4]\omega_0^{-1}$. After $6\times10^4\omega_0^{-1}$, there is the contradictory trend between the SBS of fast mode and the SBS of quasimode, which illustrates the competition of corresponding SBS of two modes. The period of the total SBS reflectivity recurrence is $T_R=7.501\times10^4\omega_0^{-1}-6.494\times10^4\omega_0^{-1}=1.007\times10^4\omega_0^{-1}$ calculated from two peaks of the SBS reflectivity as shown in Fig. \ref{Fig:Reflectivity}(a), thus the frequecy of SBS reflectivity recurrence is $\omega_R=1/T_R=9.93\times10^{-5}\omega_0^{-1}$. In the same way, the period of SBS of fast mode is $T_f=2.031\times10^4\omega_0^{-1}$ and the period of SBS of quasimode is $T_q=2.126\times10^4\omega_0^{-1}$. Thus, the sum of two periods is $\omega_f+\omega_q=1/T_f+1/T_q=9.63\times10^{-5}\omega_0^{-1}$, which is close to the period of SBS reflectivity recurrence $T_R$. This illustrates that the burst behavior of SBS reflectivity is as a result of competition between SBS of fast mode and SBS of quasimode, and the recurrence period of SBS reflectivity burst comes from the sum of two periods of SBS of fast mode and SBS of quasimode. Fig. \ref{Fig:Reflectivity}(b) demonstrates the total reflectivity and transmitivity of the SBS. Since the burst behavior of the total reflectivity of SBS will occur due to the competition between SBS of fast mode and SBS of quasimode, the transmitivity will demonstrate a same burst behavior which is complementary to the SBS reflectivity after $3\times10^4\omega_0^{-1}$. Therefore, the competition between SBS of fast mode and SBS of quasimode is an important factor on the burst behavior of SBS and the nonlinear saturation of SBS.
 
  \begin{figure}[!tp]
  	\includegraphics[width=1\columnwidth]{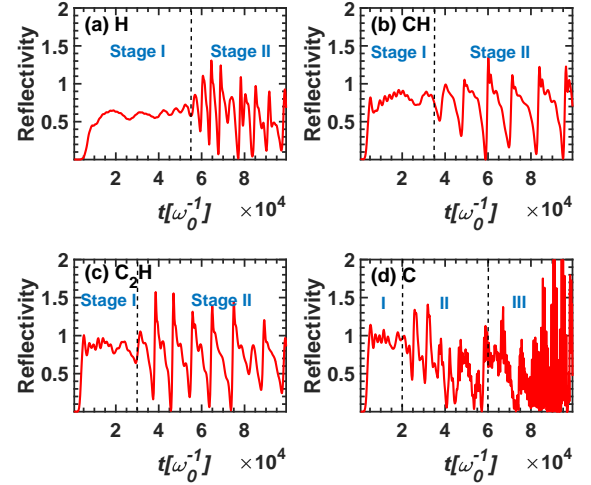}
  	\caption{\label{Fig:R_t}(Color online) The SBS reflectivities in different species plasmas evolve with time.}
  \end{figure}

Figure \ref{Fig:R_t} demonstrates the SBS reflectivities in different species plasmas. In stage I, the SBS reflectivities will develop and then saturate at a nearly fixed level. In stage II, when different modes such as the fast IAW mode and quasimode are excited in SBS in our simulation condition, other modes such as slow IAW mode, ion bulk (IBk) mode \cite{Feng_2017PPCF1} could be excited in SBS in some conditions, the competition among SBS of different modes will lead to the low-frequency burst behavior of SBS reflectivity. With the rate of C to H increasing, from Fig. \ref{Fig:R_t}(a) to Fig. \ref{Fig:R_t}(d), the Landau damping of IAW will decrease and the SBS growth rate and SBS reflectivity will increase. As a result, the beginning time of stage II will be earlier. However, for C plasmas, the Landau damping of IAW is very low and the gain of SBS is very high in the condition of $I_0=1\times10^{16}W/cm^2$, the SBS cascade will occur in stage III. \cite{Feng_2017PPCF2} The low-frequency burst in stage II is due to the competition among SBS of different modes, while the high-frequency burst in stage III is as a result of SBS cascade. These results in Fig. \ref{Fig:Reflectivity} give a clear explanation of the low-frequency burst behavior in SBS, which is common phenomena of SBS in high-intensity laser-plasma interaction.

\begin{figure}[!tp]
	\includegraphics[width=1\columnwidth]{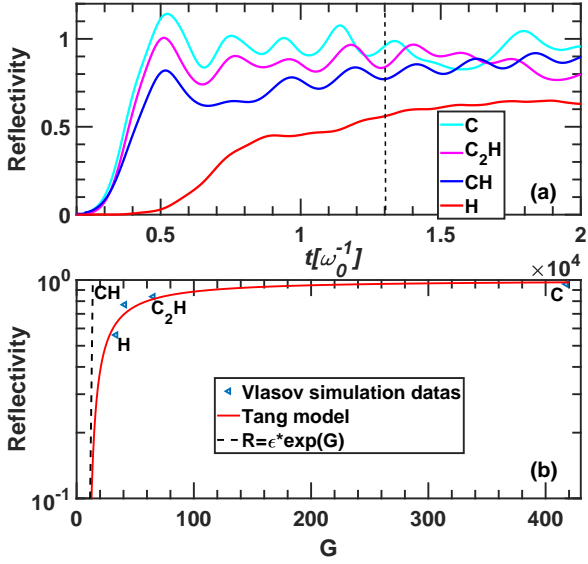}
	\caption{\label{Fig:R_G}(Color online) (a) The early linear stage of SBS in different species plasmas. (b) The relation between the SBS reflectivity and SBS gain in different species plasmas. Where the gains in multi-ion species plasmas, such as CH and C$_2$H plasmas, are calculated by the kinetic theory, and the gains in single-ion species plasmas, such as H and C plasmas, are calculated by the fluid theory. The SBS reflectivities by Vlasov simulation take the values at $t=1.3\times10^4\omega_0^{-1}$.}
\end{figure}

Figure \ref{Fig:R_G}(a) gives the linear growth and saturation process of SBS in the early stage, in which the condition is as the same as Fig. \ref{Fig:R_t}. We can see that the SBS in C plasmas increases the most quickly among the four cases, while the growth rate of SBS in H plasmas is the slowest and the saturation level is the lowest. With the rate of H to C in the plasmas increasing, the growth rate and saturation level will decrease obviously, which is because of increase of the IAW Landau damping. According to Fig. \ref{Fig:R_G}(a), the SBS reflectivity values at the time of $t=1.3\times10^{4}\omega_0^{-1}$ are chosen as the saturation values, which is shown in Fig. \ref{Fig:R_G}(b), because they are stable at this time. The linear gain of SBS by fluid theory is given by
\begin{equation}
\label{Eq:G_B}
G=2\frac{\gamma_{0B}^2}{\nu_Av_{gs}}L,
\end{equation}
where 
\begin{equation}
\gamma_{0B}=\frac{1}{4}\sqrt{\frac{n_e}{n_c}}\frac{v_0}{v_{te}}\sqrt{\omega_0\omega_A}
\end{equation}  
is the maximum temporal growth rate of SBS \cite{Berger_2015PRE,Lindl_2004POP},  $v_0=eA_0/m_ec$ is the electron quiver velocity. $v_{gs}=c^2k_s/\omega_s$ is the group velocity of SBS scattering light, $L$ is the plasmas density scale length, $\omega_A\equiv Re(\omega_A)$ and $\nu_A\equiv |Im(\omega_A)|$ are the frequency and Landau damping of IAW. The collision damping of IAW can be neglected since the electron temperature is as high as $T_e=5keV$ in our simulation, thus only the Landau damping of IAW is considered. For single-ion species plasmas, such as H or C plasmas, the gain of SBS can be calculated by the fluid theory. However, for multi-ion species homogeneous plasmas, the SBS gain can be calculated by the kinetic theory \cite{Lindl_2004POP}:
\begin{equation}
G(\omega_s)=\frac{1}{4}\frac{k_A^2v_0^2L}{v_{gs}\omega_s}\text{Im}(\frac{\chi_e(1+\sum\limits_{i}\chi_i)}{\epsilon(k_s-k_0, \omega_s-\omega_0)}),
\end{equation}
 which is more precise than fluid theory. Where subscripts $0, s, A$ represent the pump light, SBS scattering light and IAW.
Under the strong damping condition $\nu_A/\gamma_{0B}*\sqrt{v_{gs}/v_{gA}}\gg1$ \cite{Forslund_1975POF}, one can get the Tang model \cite{Tang_1966JAP}:
\begin{equation}
R(1-R)=\varepsilon\{\text{exp}[G(1-R)]-R\},
\end{equation}
where $R$ is the reflectivity of SBS at the left boundary, and $\varepsilon$ is seed light at the right boundary. If $R\ll1$, the Tang model can be approximate to the seed amplification equation:
\begin{equation}
\label{Eq:seed amplification}
R=\varepsilon\cdot\text{exp}(G).
\end{equation}
   
Although the strong damping condition in C plasmas or C$_2$H plasmas is not satisfied, the Tang model can also be applied to predicting the SBS reflectivity in the linear saturation stage. As shown in Fig. \ref{Fig:R_G}(b), the SBS reflectivities in different species plasmas are close to the Tang model. And the seed amplification equation can not be applied to predicting the SBS reflectivities in the condition of strong pump light, since the SBS reflectivity $R$ is not much lower than 1. Figure \ref{Fig:R_G} shows the linear process, including linear growth and saturation process, in which the linear theory such as Tang model is applicable. However, in the nonlinear process, such as stage II and stage III in Fig. \ref{Fig:R_t}, the burst behavior will occur and the nonlinear saturation mechanism of SBS has been explained in this paper.

%\section{\label{Sec:Summary}Summary}
In conclusions, quasimode will be excited by SBS in high-intensity laser-plasma interaction, which will compete with the IAW excited by SBS. The competition between SBS of quasimode and SBS of fast mode in C$_2$H plasmas demonstrates that this competition mechanism is the cause of the low-frequency burst behavior of SBS. In different species plasmas, the same low-frequency burst behavior will occur, which illustrates that the competition mechanism is common in high-intensity laser-plasma interaction no matter what the plasmas are. These results also give a good explanation of the intermediate low-frequency burst and saturation process in C plasmas.

%\begin{acknowledgments}
We would like to acknowledge useful discussions with C. Z. Xiao and L. Hao. This research was supported by the National Natural Science Foundation of China (Grant Grant Nos. 11875091, 11575035, 11475030 and 11435011), National Postdoctoral Program for Innovative Talents (No. BX20180055), the China Postdoctoral Science Foundation (Grant No. 2018M641274) and Science Challenge Project, No. TZ2016005.
%\end{acknowledgments}

%\bibliographystyle{apsrev4-1}
%\bibliographystyle{aipnum4-1}
\bibliography{SBS_multi_species.bib}

\end{document}